\documentclass[journal]{IEEEtran}
\usepackage{cite}

\ifCLASSINFOpdf
   \usepackage[pdftex]{graphicx}
   \graphicspath{{./img/}{./}}
   \DeclareGraphicsExtensions{.pdf,.jpeg,.png}
\else
\fi

\usepackage{amsmath}

\begin{document}
%
% paper title
% Titles are generally capitalized except for words such as a, an, and, as,
% at, but, by, for, in, nor, of, on, or, the, to and up, which are usually
% not capitalized unless they are the first or last word of the title.
% Linebreaks \\ can be used within to get better formatting as desired.
% Do not put math or special symbols in the title.
\title{Low latency communication over commercially available LTE and remote driving}

% author names and affiliations
% transmag papers use the long conference author name format.

\author{\IEEEauthorblockN{Andrey Belogolovy, Deepak Dasalukunte, Richard Dorrance, Evgeny Stupachenko, and Xue Zhang}
\thanks{The authors are with Intel Corp. Corresponding author: Andrey Belogolovy (email: andrey.belogolovy@intel.com)}}

% The paper headers
%\markboth{Transactions on \LaTeX\ Vehicular Technology}%
% The only time the second header will appear is for the odd numbered pages
% after the title page when using the twoside option.
% 
% *** Note that you probably will NOT want to include the author's ***
% *** name in the headers of peer review papers.                   ***
% You can use \ifCLASSOPTIONpeerreview for conditional compilation here if
% you desire.

% If you want to put a publisher's ID mark on the page you can do it like
% this:
%\IEEEpubid{0000--0000/00\$00.00~\copyright~2015 IEEE}
% Remember, if you use this you must call \IEEEpubidadjcol in the second
% column for its text to clear the IEEEpubid mark.

% use for special paper notices
%\IEEEspecialpapernotice{(Invited Paper)}

% for Transactions on Magnetics papers, we must declare the abstract and
% index terms PRIOR to the title within the \IEEEtitleabstractindextext
% IEEEtran command as these need to go into the title area created by
% \maketitle.
% As a general rule, do not put math, special symbols or citations
% in the abstract or keywords.
\IEEEtitleabstractindextext{%
\begin{abstract}

%One of the emerging applications for the Ultra Reliable and Low Latency Communication (URLLC) is Vehicle to Everything (V2X). V2X can augment the onboard sensors for the autonomous driving, specifically the concept of platooning where the vehicle train follow a lead vehicle.  Sometimes it is desirable to drive the lead vehicle remotely, and in this case, it is very critical to have a predictable latency for remote navigation.  In this paper, we present a communication and video streaming system that addresses the challenges and enables teleoperation of a real car.

%In addition to autonomous car operation, in many cases it is desirable to let a human drive the vehicle remotely. To make remote operation possible, it is very critical to have a low and predictable latency to transmit video from the car cameras and receive control commands back. In this paper, we analyze the problem and present a communication and video streaming system that addresses the latency challenges and enables teleoperation of a real car over commertially available LTE network.

In addition to autonomous car operation, in many cases it is desirable to let a human drive the vehicle remotely. To make remote operation possible, it is very critical to have a low and predictable latency to transmit video from the car cameras and receive control commands back. In this paper, we analyze the problem and present a communication and video streaming system that addresses the latency challenges and enables teleoperation of a real car over commercially available LTE network; demonstrating sub-50ms roundtrip latencies for 720p, 60FPS video, with average PSNR 36db.

\end{abstract}

% Note that keywords are not normally used for peerreview papers.
%\begin{IEEEkeywords}
%IEEE, IEEEtran, IEEE Transactions on Magnetics, journal, \LaTeX, magnetics, paper, template.
%\end{IEEEkeywords}
}

% make the title area
\maketitle

% To allow for easy dual compilation without having to reenter the
% abstract/keywords data, the \IEEEtitleabstractindextext text will
% not be used in maketitle, but will appear (i.e., to be "transported")
% here as \IEEEdisplaynontitleabstractindextext when the compsoc 
% or transmag modes are not selected <OR> if conference mode is selected 
% - because all conference papers position the abstract like regular
% papers do.
\IEEEdisplaynontitleabstractindextext
% \IEEEdisplaynontitleabstractindextext has no effect when using
% compsoc or transmag under a non-conference mode.

% For peer review papers, you can put extra information on the cover
% page as needed:
% \ifCLASSOPTIONpeerreview
% \begin{center} \bfseries EDICS Category: 3-BBND \end{center}
% \fi
%
% For peerreview papers, this IEEEtran command inserts a page break and
% creates the second title. It will be ignored for other modes.
\IEEEpeerreviewmaketitle

\section{Introduction}
\label{sec:intro}
\begin{figure}[t]
\centering
\includegraphics[
width=\columnwidth
]%
{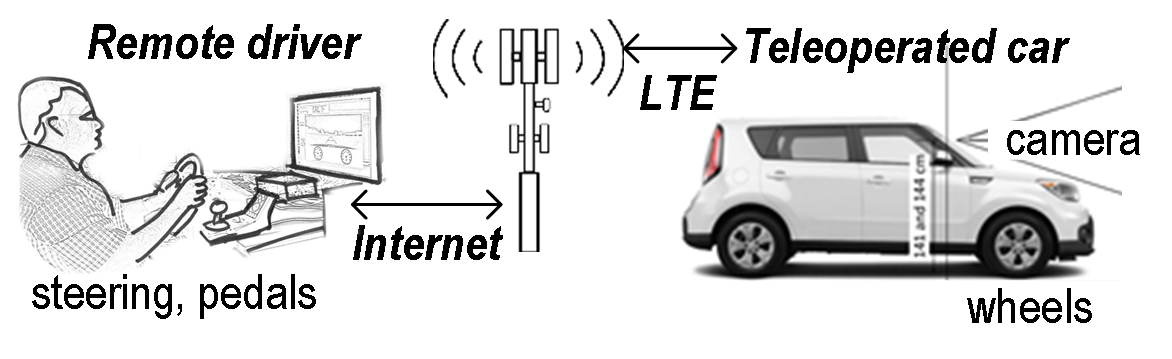}
\caption{Remote driving framework.}
\label{Fig:framework}
\end{figure}

\IEEEPARstart{R}{emote} driving is a term used to describe the teleoperation of a car where the driver is physically located far away from the car, but can see video from a windscreen mounted camera, turn the steering wheel, and press the pedals to behave like a real driver, with steering, braking and acceleration commands sent to the car as shown in Figure \ref{Fig:framework}. 

Today remote driving is considered as a companion to almost all autonomous driving system to increase overall car operation reliability and safety. For example, in the case of an unexpected event—like a road work enforced lane change, flood, fire, emergency or similar—when the autonomous driving system does not know how to react, it requests a remote driver to connect to the car, evaluate the situation, and continue driving; instead of stopping and making an autonomously driven car a problem to other traffic parties \cite{Kang18}.

To enable the remote driving, many technical problems should be solved: a car needs to have Drive-By-Wire (DbW) capability that assumes electronic, not mechanical control of  steering, braking, and other systems; a system to transmit video and other measurement data from the car to the remote operator; and a similar system to send and receive commands from the remote operator. 

Latency is a key factor in DbW control, video compression, and data transmission and reception. However, in the literature the authors did not find exact numerical parameters (like average and variance) that are required to say that some number is a maximum, i.e., if a parameter of the system is greater than some value, then the usage is not possible. We decided to build a prototype to answer the question with practical tests and measurements similar to \cite{Luck06} where communication latency acceptability was studied with the real experiment of robot teleoperation.

The end-to-end latency of any teleoperation, including remote driving, is more than just the network packet roundtrip time. The true end-to-end operating latency is a sum of all individual framework component latencies, including low-level processing and buffering. There is no single hotspot that can be optimized to reduce the overall latency dramatically as there are many components that are either out of an engineer’s control or very hard to modify. For example, displaying framebuffer is a source of latency, just as any other block-based processing element in the system, since it waits for the whole block to be received before continuing. Table \ref{tab:table1} shows typical latency sources in the system in the same order as the video is transmitted from the source to the remote driver camera capture, framebuffer and display refresh are connected to frame rate, 60fps in this case. Video compression and de-compression, LTE and internet communication links introduce variable latency and are as indicated in Table \ref{tab:table1}.

\begin{table}[h!]
\begin{center}
\caption{Sources of Latency in Remote Driving}
\label{tab:table1}
\begin{tabular}{c|c|c|}
Source &\vtop{\hbox{\strut Latency}\hbox{\strut [ms]}} &\vtop{\hbox{\strut Cumulative Latency}\hbox{\strut [ms]}}\\
\hline
Camera capture, 60fps & 17 & 17\\
Video compression & 17 \ldots 120 & 34 \ldots 137 \\
\vtop{\hbox{\strut LTE+Internet one way,}\hbox{\strut 1.5K packet}} & 10 \ldots 300 & 44 \ldots 437 \\
Video decompression & 17 \ldots 32 & 61 \ldots 469 \\
Framebuffer & 17 & 78 \ldots 486 \\
DVI, LCD refresh & 10 & 88 \ldots 496 \\
\end{tabular}
\end{center}
\end{table}

In this paper, we present a way to reduce the end-to-end latency significantly, so even with existing LTE mobile networks it would be possible to enable latency critical workloads like remote teleoperation of a real car without any special agreement from cellular carriers, such as free traffic or prioritization of data. We present a remote driving framework in which we tackled two main latency sources:

\begin{enumerate}
\item Network latency and buffering;
\item Video encode and buffering.
\end{enumerate}
The paper is organized as follows: in Section \ref{sec:2}, we review the theory of the network latency reduction with forward error correction coding \cite{Kabatiansky05} and present hypothetical estimations of gains. However, to enable the theory, many implementation considerations are discussed in Section \ref{sec:PRACTICAL}. In  Section \ref{sec:slice}, we describe a video compression scheme and its practical implementation for remote car operation. Section \ref{sec:summary} summarizes the results.

\section{Forward error correction application at the network layer to reduce transmission latency }
\label{sec:2}
To send a large data block (like a video slice), the application has to split data into K smaller packets in accordance to the physical, MAC and higher layer protocols. Then at the receiver side, the whole data block can be processed only after all of its contents are received. So, the transmission latency is determined by the latest packet of the original block that arrived. Network layer coding is a well-known theory to reduce delay in packet switching networks \cite{Kabatiansky05} by introducing redundancy: original $K$ packets are considered as symbols of some erasure correction code (ECC) and encoded into $N > K$ symbols (packets), then all $N$ packers are sent to the network. Usually, in accordance to the properties of the ECC, the decoder can reconstruct the codeword by any $K$ out of $N$ packets, so the latency of the whole block will be determined by the latest out of $K$ first received packets. ($k$-order statistic). To demonstrate the effect of the network layer coding in terms of reconstructed block latency distribution, consider a theoretical example below.

\begin{figure}[t]
\centering
\includegraphics[
width=\columnwidth
]%
{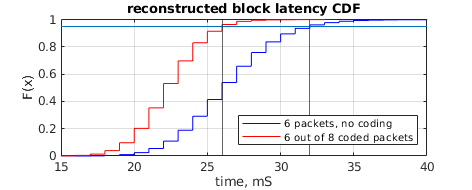}
\caption{An example of theoretical latency gain of network coding}
\label{Fig:example}
\end{figure}

Assume we transmit data blocks that will be split into $10$ packets, and the packet delay time are independent Gaussian random numbers with mean $=20$ and stddev $=5$. Original reconstructed block latency can be drawn as a cumulative distribution function (CDF) of maximum of $6$ samples. If we encode $6$ packets into $8$ and assume that the network load did not change, so the packet delay time has the same distribution, we can plot a $6$-order statistic for $8$ samples, and it will represent a theoretical latency of the reconstructed blocks with the network layer coding (code rate $=3/4$), see Figure \ref{Fig:example}. At the level of probability $=0.95$ one can see the difference of $6$. So, if random numbers represent milliseconds of a packet delay, in the given example the network coding can provide $6$ms latency improvement.

The numbers look so great so many companies wanted to use network layer coding for latency critical applications like voice/video conferencing, control systems and other and-user applications over the Internet. But in practice, end-user applications are different than the ideal model because of the following:

%\begin{enumerate}
%\item 
1. Erasure correction codes used in applications today are not optimal. In 1985, when Kabatiansky and Krouk first published their paper that coding reduces latency, the computational complexity of encoding/decoding was a big concern, so many other researchers proposed code constructions like LT codes, Fountain and Raptor codes \cite{Luby07} that provided linear decoding complexity. They are very implementation friendly, but unlike Reed-Solomon codes that are optimal in terms of capacity, Raptor codes are not. 

\begin{figure}[t]
\centering
\includegraphics[
width=\columnwidth
]%
{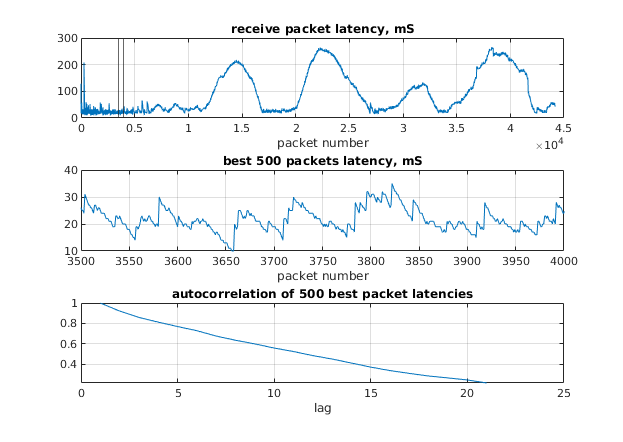}
\caption{Latency of a constant rate stream measured at the receiver and autocorrelation of $500$ best}%
\label{Fig:constantRate}
\end{figure}

%\item 
2. Data buffers in modems and other network interfaces. When an application opens an IP socket and sends a datagram, it does not mean that the packet is actually sent to the physical layer of the network. There are many buffers at the operating system level also as at the level of the device, so totally some number of megabytes can be buffered before physical transmission. Figure \ref{Fig:constantRate} presents an example of a packet latency measured at the receiver for $1$MBps constant transmission rate where one can see that at some time latency values grow to $200ms$ meaning that the packets were just buffered somewhere in the transmission pipeline when the channel rate was low, and when the network conditions changed, packets started to arrive.

\begin{figure}[t]
\centering
\includegraphics[
width=\columnwidth
]%
{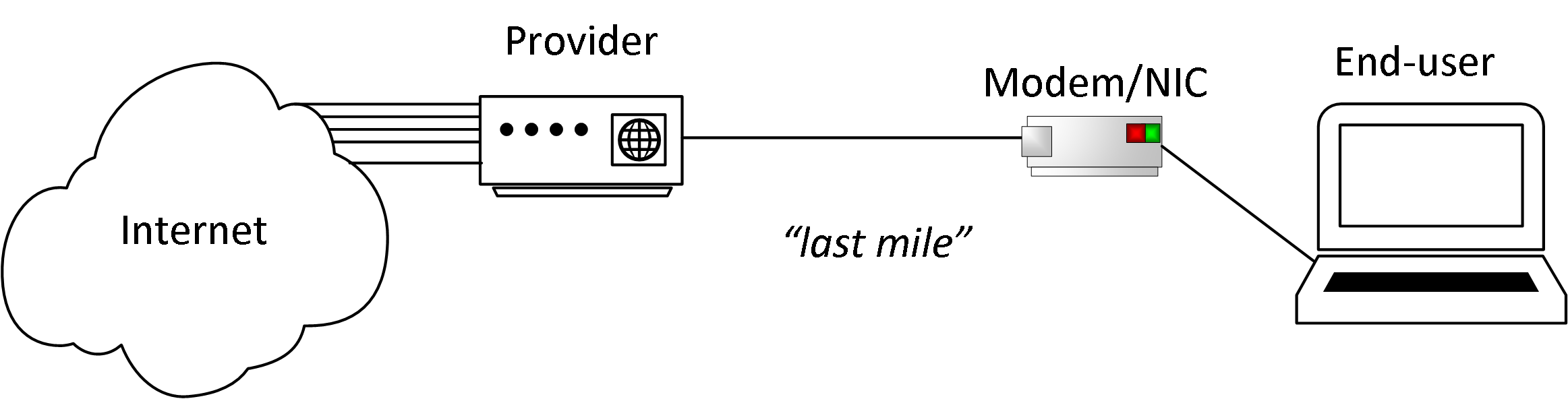}
\caption{A typical end user connected to the Internet with a single link.}
\label{Fig:singleModem}
\end{figure}

%\item 
3. Packet delay times are not independent. The Internet itself is a multi-hop packet switching network at the level of switches and routers, but when it comes to the last mile, a typical user has only one line of connection like shown in Figure \ref{Fig:singleModem}, and this last mile makes packet delay times very correlated: if the last mile packet is delayed, the probability that the next packet will be also delayed is very high as there is only one physical link, and packets are sent one-by-one into it. In Figure \ref{Fig:constantRate} the graph at the bottom shows autocorrelation of $500$ best packet latencies, and it is very high. 

%\end{enumerate}

\section{Practical aspects of low latency transmission for commercially available LTE}
\label{sec:PRACTICAL}
In this section we describe practical methods of solving three problems mentioned above to enable latency reduction with network coding.

\subsection{High-throughput implementation of optimal erasure correction codes}
At the time of Raptor codes appearance, Reed-Solomon based erasure control codes were considered as not implementable to provide data rates at the order of number of megabytes per second and above. But since that time, computer hardware and instruction set architectures has made significant progress, and today it is not a problem to reach hundreds of megabytes per second with Cauchy Reed-Solomon codes (CRS) using SSE2 \cite{Taylor}:

The target data rates for LTE modems we planned to use is in the order of 1-2MBps, we have chosen CRS codes as they always provide required data rates and are optimal from the redundancy point of view. Table \ref{tab:CRS} lists the CRS performance for a packet size of 1296 bytes under various redundancy levels.

\begin{table}[h!]
\begin{center}
\caption{CRS performance for packet length 1296 bytes}
\label{tab:CRS}
\begin{tabular}{c|c|c|}
\vtop{\hbox{\strut Code message length K }\hbox{\strut and number of } \hbox{\strut redundant symbols M}} &\vtop{\hbox{\strut Encoder }\hbox{\strut [Mbps]}} & \vtop{\hbox{\strut Decoder  }\hbox{\strut [Mbps]}}\\
\hline
K=100, M=1 & 23314 & 19259\\
K=100, M=10 & 1171 & 1181 \\
K=100, M=20 & 579 & 618 \\
K=100, M=30 & 347 & 401 \\
K=100, M=40 & 274 & 298 \\
K=100, M=50 & 218 & 237 \\
\end{tabular}
\end{center}
\end{table}

\subsection{Instant rate control and scheduling make possible to avoid buffering}
To avoid packet buffering at the transmitter side, we used a feedback-based rate control that measures the data rate at the receiver, makes a decision about the best rate for the current channel state and sends this information to the transmitter. Because we deal with video, it is always possible to increase or decrease compression ratio to match the channel rate requirements set by the rate control.
There are 2 states of the channel rate control algorithm:
\begin{enumerate}
\item Measurements and decision making. In this mode, if the measured data rate is less than the target, the target rate is set to the measured minus some delta to empty buffers if they have extra packets.
\item Rate probation. The algorithm enters probation state if there were no rate decreases for some number of measurements. In probation state, the algorithm requests temporary rate increase by some delta followed by the rate decrease by the same delta, and then measures instant data rates at the receiver. If measurements show the same increased and decreased rate, then the algorithm sends a permanent rate increase command, otherwise the rate remains unchanged.
\end{enumerate}
As we will have multiple links, we need a scheduler mechanism to distribute data packets over multiple transmitters. For each link, the scheduler intoduces a parameter we refer to as an anticipated transmission end time, and each time a new packet is given to the scueduler, the packet is assigned to the link with the closest transmission end time.
At the beginning, the anticipated transmission end time parameter is set to the current time. Then the scheduler counts the amount of data to be sent over the link, also takes measurements from the rate control to determine the expected time of transmission by simply dividing the size by the rate and increases the anticipated transmission end time by the expected time of transmission. 

\subsection{Last mile connection with multiple LTE modems}
\begin{figure}[t]
\centering
\includegraphics[
width=\columnwidth
]%
{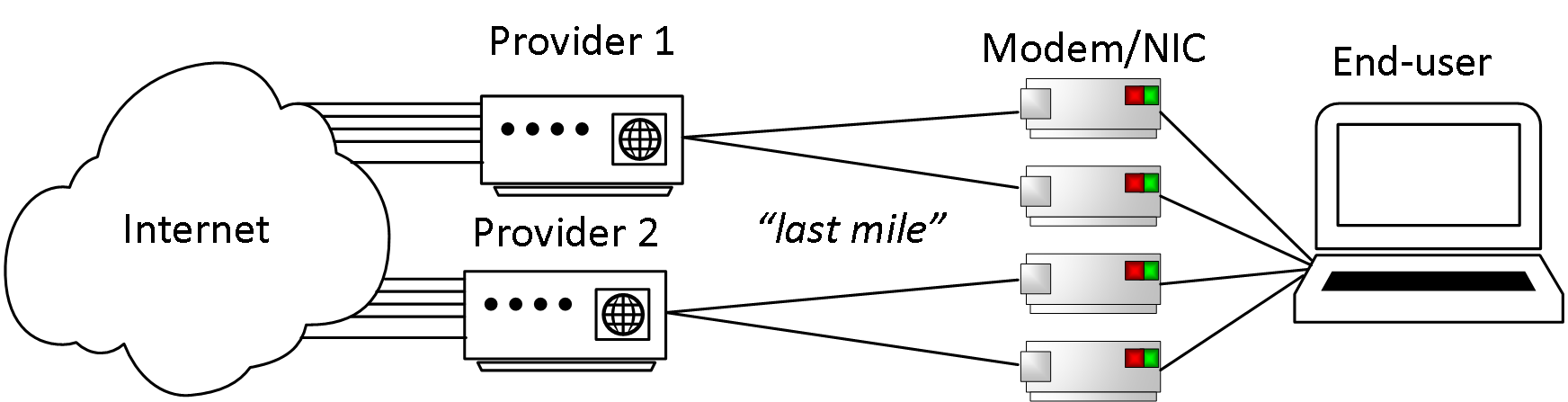}
\caption{Configuration with multiple modems from different carriers.}
\label{Fig:multipleModems}
\end{figure}

\begin{figure}[t]
\centering
\includegraphics[
width=\columnwidth
]%
{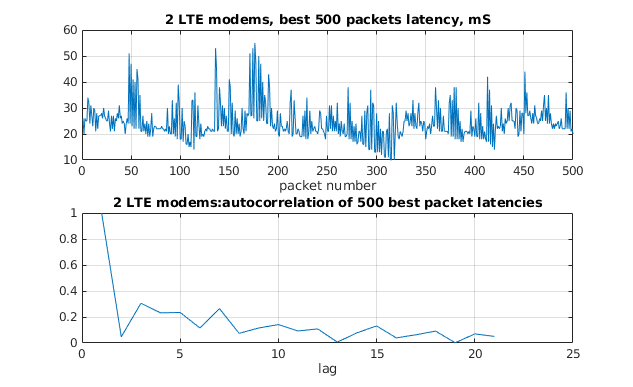}
\caption{Latency and autocorrelation of packet latency of a constant rate stream transmitted over $2$ LTE modems.}
\label{Fig:lat2}
\end{figure}

We used an extensive way to solve the last mile packet delay time correlation problem: instead of just one LTE modem, we used $4$ modems of the same type with $4$ SIM cards ($2$ SIMs from one carrier and $2$ SIMs from another carrier) as shown in Figure \ref{Fig:multipleModems}. Experiments show that even using $2$ modems the latency autocorrelation function decreases significantly (see Figure \ref{Fig:lat2}), and if the number of modems becomes $4$, there is no correlation in packet latency at all. 

\begin{figure}[t]
\centering
\includegraphics[
width=\columnwidth
]%
{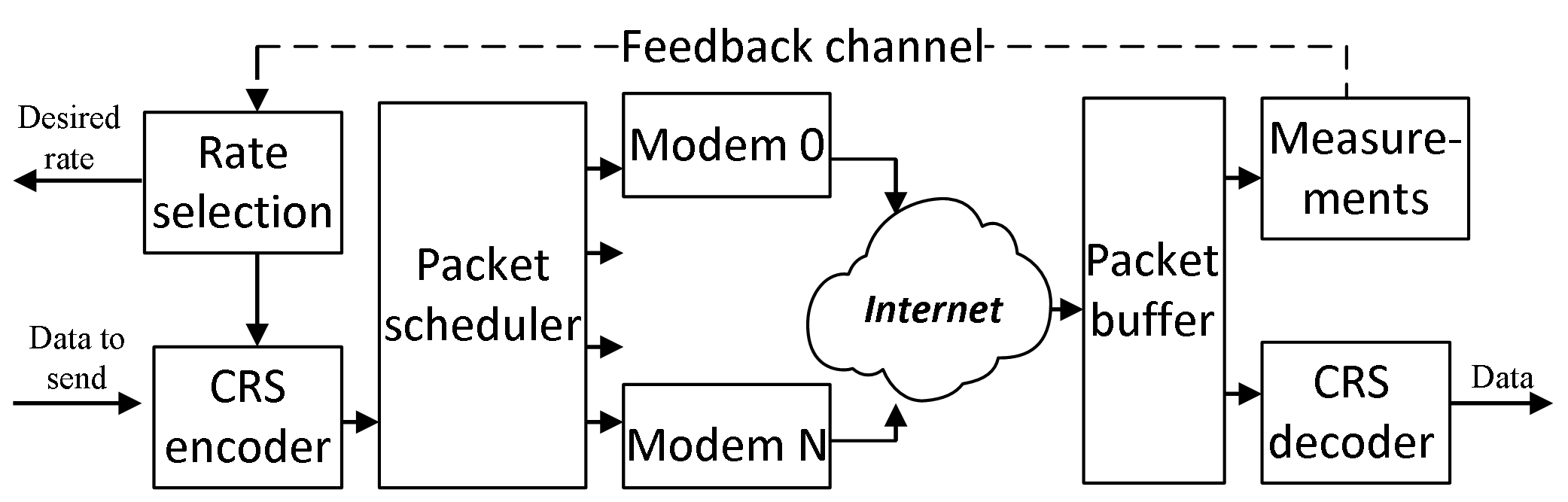}
\caption{Networking module scheme.}
\label{Fig:networking}
\end{figure}

\begin{figure}[t]%
\centering
\includegraphics[
width=\columnwidth
]%
{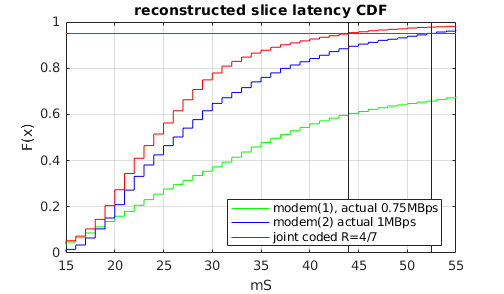}%
\caption{Reconstructed slice latency for $2$ independent modems and joint network coded at the same data rate.}
\label{Fig:joint_coded}%
\end{figure}

After implementation of the rate control, scheduling and network coding described above, we have a networking module depicted in Figure \ref{Fig:networking}. This module was using $2$ modems (SIMs were from $2$ different carriers) using the following scenario: we tested average data rates that we can have with the rate control for each link and measured packet latency. These rates were approximately $1$ and $0.75$Mbps. Then we generated a stream of data blocks with the rate of $1$MBps and encoded it with the code at rate $4/7$ so the resulting data rate was approx. $1.75$MBps that is a sum of the individual modem rates. Then we transmitted encoded packets over the same network at the same place, used FEC decoding at the receiver to reconstruct blocks right after we had enough packets and measured the latency. The latency CDF is shown in Figure \ref{Fig:joint_coded}. Curves ``modem ($1$)'' and ``modem ($2$)'' represent a hypothetic reconstructed latency of the same block without coding using the latency statistics measured at the beginning of the experiment.  At the level of probability $=0.95$ we had latency gain of $8.5$ms.

\section{Slice based video compression system}
\label{sec:slice}

A traditional $60$fps camera outputs video at the frame granularity, i.e. every $\left(\frac{1}{60}\right)^{\text{th}}$ of a second it outputs a data block corresponding to the whole frame. There is no other way to make the latency lower but to reduce the granularity. At the time of making the prototype, we did not find any cameras in the market that were able to output data at a granularity lower than one frame, so we developed a dedicated system with the following requirements:
\begin{enumerate}
\item Read a slice (a horizontal part of the frame) from the image sensor and pass it to the compression. We have chosen to read data with the granularity of $\frac{1}{6}$ of the frame at $60$fps, so the read latency = $\frac{6}{60} = 0.002(7)$sec.;

\item Get the current available rate from the networking rate control and scheduler module and compress image slice to be transmitted in accordance with the compression ratio derived from the desired data rate.
\end{enumerate}

\subsection{H.264 intra refresh compression mode for streaming}

\begin{figure}[t]%
\centering
\includegraphics[
width=\columnwidth
]%
{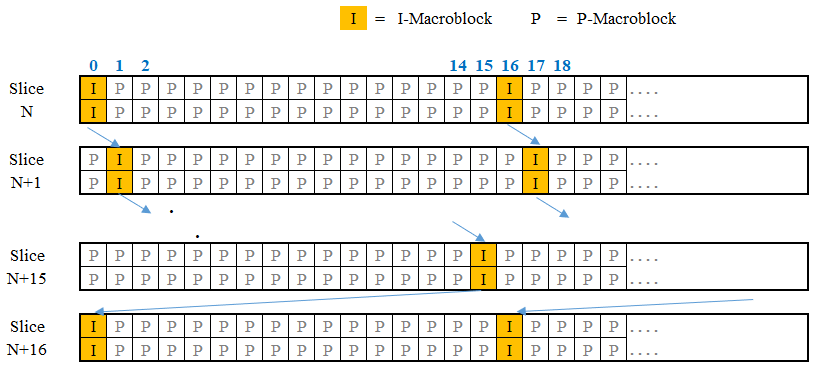}%
\caption{Sweeping I-macroblock pattern in a P-slice.}
\label{Fig:slidingIPcolumn}%
\end{figure}

We have chosen H.264 \cite{Wiegand03} as a video compression option as it provides high compression ratios with a various number of features to recover from the packet loss recovery at the time of transmission, also there were many available libraries to decode H.264 streams in software. By streaming mode, we assume that at the receiver side the video frame needs to be displayed at a certain time no matter whether the decoder has all frame slices available or not. In other terms, no buffering, no waiting for the slices that are either delayed or lost in transmission. As we wanted to minimize the video display latency, we chose the display time close to transmission reconstruction time at $95\%$ probability, so we had a non-zero probability of slice loss at the input of the video decompressor that can be roughly approximated in a following way: one slice out of $20$ is lost. In this case, lossy reconstruction features of the H.264 standard became very important. To operate in the case of slice loss,  we used an option in H.264 called “intra refresh”, when intra-predicted macroblocks are systematically placed in the inter-predicted slices instead of sending entire intra frames, see Figure \ref{Fig:slidingIPcolumn}.

\begin{figure}[t]%
\centering
\includegraphics[
width=\columnwidth
]%
{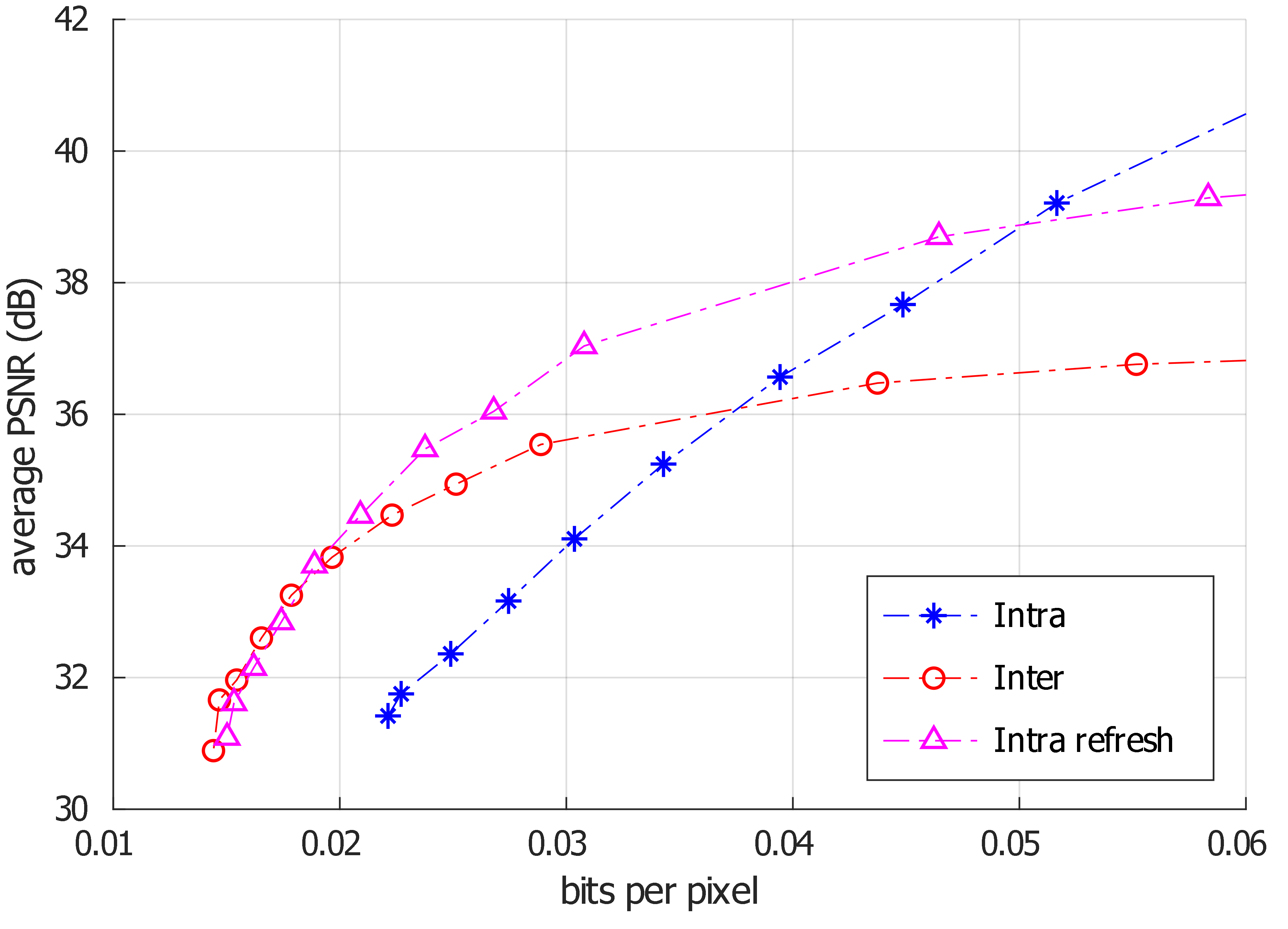}%
\caption{Rate-distortion graph for Intra, Inter and Intra-refresh.}
\label{Fig:rateDis}%
\end{figure}

In our scheme, we have chosen columns of I-macroblocks spaced every $16$ macroblocks as the granularity of intra refresh; at any slice $N$, if the I-macroblock columns appear at index $0, 16, 32, \ldots$ the subsequent slice $(N+1)$ has I-macroblock columns at indices $1, 17, 33, \ldots$. At slice $N+16$, the I-macroblock columns restart from indices same as that of slice $N$. With this mechanism, even with some dropped frames in between, recovery of the image happens within $16$ frames. In a camera with $60$fps frame capture rate, the recovery time will be $0.26$s which is same as the average reaction time of a human to visual stimulus. The recovery time can be adjusted based on the repetition rate of the I-macroblocks at the cost of slightly higher data rate requirement. I-macroblocks with a periodicity of $8$ reduces the recovery time by $2$x (to $0.13$s) and so on. A recovery rate of $\leq 0.26$s suits perfectly well to requirements such that any dropped frames can be recovered without the remote driver/operator noticing any change in the video being transmitted. Besides just visual evaluation, we have studied the effect of intra-refresh in the case of packet loss comparing $3$ modes of the H.264 encoder that we have implemented: intra-only mode (all frames are encoded as intra, so they don’t require reference frames to be decompressed), inter-only mode (only the first frame in the stream is encoded as intra, and all others are encoded as inter referring always to the previous frame) and intra-refresh mode (described above). We present PSNR/rate graph for image sequences of $16$ slices captured with the sensor mounted on the car windscreen and is presented in Figure \ref{Fig:rateDis}. Here we intentionally dropped one compressed slice, and then decompressed the rest $15$ with FFMPEG \cite{FFMPEG} software and measured average PSNR over the decompressed slices assuming that the first missing slice was directly duplicated from the slice previous to the dropped one. As one can see from the graph, the encoder in intra-refresh mode outperforms others at the most important PSNR range (about $36-37$dB), so we have proven that intra-refresh is the best option.

\subsection{Real time quantizer update}

\begin{figure}[t]%
\centering
\includegraphics[
width=\columnwidth
]%
{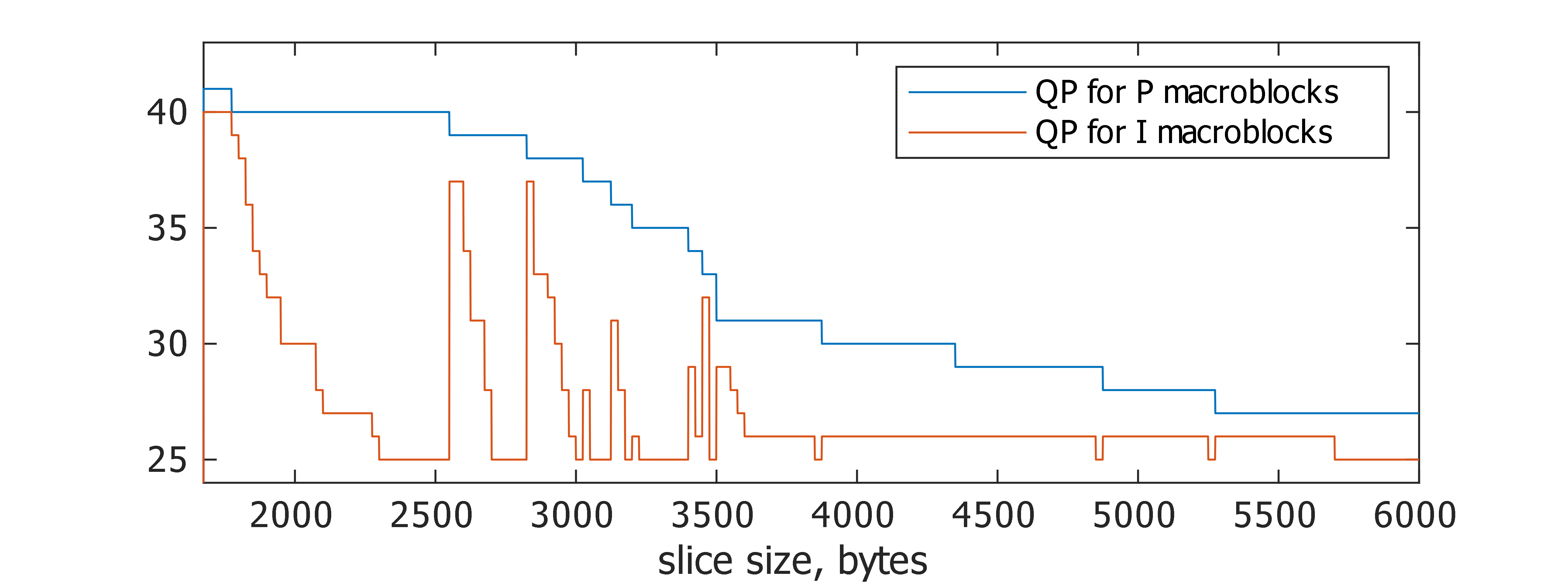}%
\caption{Graphical representation of the lookup table for quantizer selection as a function of a desired compressed slice size.}
\label{Fig:graphical}%
\end{figure}

H.264 standard provides developers with a lot of options affecting compression ratio, but for simplicity, in our system we used only quantization parameters update for I and P macroblocks to change the size of the compressed slices. As our target was low latency, we implemented a simple lookup-table based quantizer selection procedure based on statistical data. In the video compression unit, we implemented cyclic sequential increase and reset of the quantization parameters, then a sensor mounted to the windscreen of the real car was driven on the road for 1 hour to collect statistics of compressed slice length as a function of I and P macroblock quantizers. The compression was done using intra-refresh option presented in the previous subsection. Then we averaged the statistics collected and built a look-up table where for each desired slice size had a corresponding pair of quantizers (for I and P macroblocks) to be used. A graphical representation of the table  depicting the QP values for desired compressed slice sizes is presented in Figure \ref{Fig:graphical}.

\subsection{Tightly coupled image sensor and ping-pong video encoder architecture}
\begin{figure}[t]%
\centering
\includegraphics[
width=\columnwidth
]%
{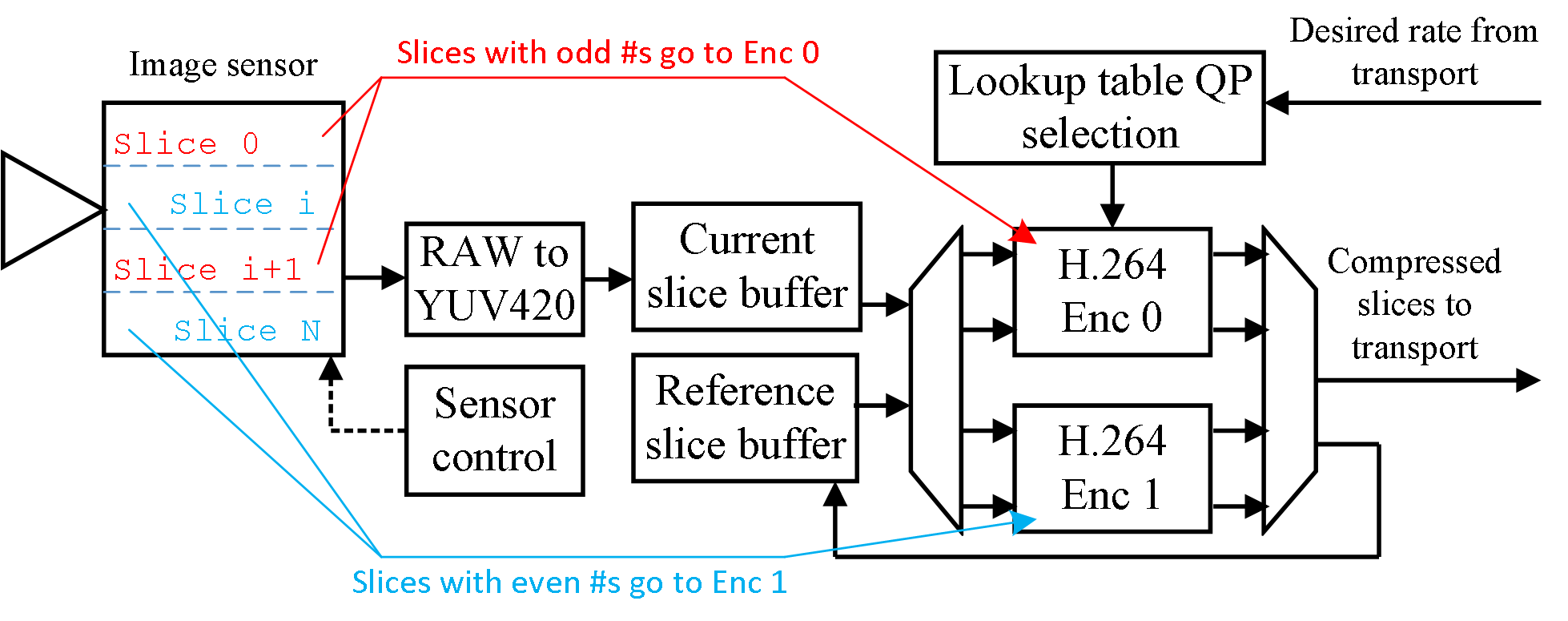}%
\caption{Video compression architecture.}
\label{Fig:video}%
\end{figure}

H.264 encoder requires the whole slice to be available in its memory before starting compression because of predictions. Even in the case of reading data from the sensor directly, data needs to be buffered. Another important feature of the H.264 compression is that its entropy encoding part operation time is data dependent, i.e. it is not possible to predict the exact number of operations that CABAC or CAVLC encoder will require in advance. In order not to waste the sensor buffering time, and in order to have a guard time interval for the entropy encoder to complete all its operations, instead of a single H.264 encoder, we decided to use $2$ independent instances of the encoders. The architecture of using $2$ encoders is shown in Figure \ref{Fig:video}; after reading image data from the sensor, it is converted to YUV420 format on the fly and stored in the local memory (slice buffer), and the first instance of the encoder starts to process the buffered video slice. At the same time, when the first encoder is processing the slice, sensor data reading and YUV420 conversion is continues to prepare the next slice writing it into which is eventually processed by the second instance of the H.264 encoder. The sensor continuously reads the image data and writes the slices into alternating buffers which are in turn processed by the encoders in a ping pong fashion. As a result, such a ping-pong scheme and $2$ instances of H.264 encoder doubles the slice reading time to complete all operations. In the case of $6$ slices per frame, each encoder has $\frac{6\times 2}{60} = 0.00(5)$ seconds, that was our time boundary for the hardware. The overall latency of the sensor-compression pair was approx. $8.3$ms.
 
\begin{figure}[t]%
\centering
\includegraphics[
width=\columnwidth
]%
{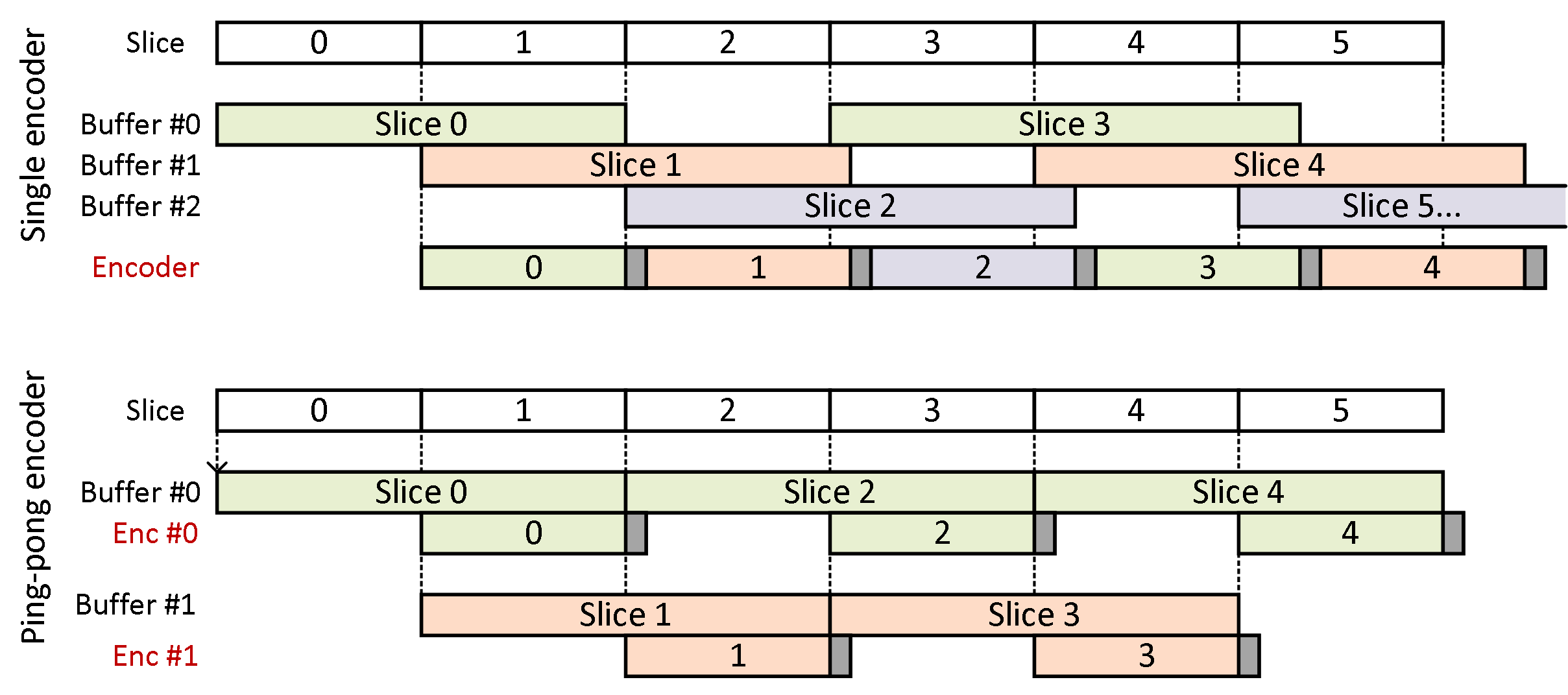}%
\caption{Timing diagram of conventional vs. ping-pong mechanism.}
\label{Fig:timing}%
\end{figure}

Figure \ref{Fig:timing} provides an approximate timing diagram of buffer usage and H.264 encoder processing for two scenarios: (1) single encoder instance; (2) $2$ instances of H.264 in a ping-pong scheme as described above. We will show that the ping-pong scheme is optimal from buffering and latency point of view. The top half of the figure shows the timing when a single encoder instance is used. The pixels from the video slice are buffered into slice buffers. The H.264 encoder can start only after the availability of slice $0$ data. Since the processing exceeds the time interval of the next incoming slice, the new slice is buffered into buffer $\#1$. As Slice $0$ is still being processed, slice $2$ data has to be buffered into buffer $\#3$. The encoder sequentially processes each slice while the buffer requirement is at $3$ times the size of one video slice.

Now by doubling the H.264 encoders, we see that (from second half of Figure \ref{Fig:timing}), there is a buffer associated with each encoder and they process either the even or odd numbered slices. Thus, the buffer requirement is reduced compared to when using a single encoder instance and at the same time halving the latency. Hence the ping-pong scheme improves the latency contributing to the overall reduction in latency of the entire system. 

The compression scheme is connected to the networking module: after receiving the desired rate, quantization parameters are chosen using the look-up table described in the previous subsection and used in the compression. The compressed slices are sent to the networking module for transmission.  

\section{Summary}
\label{sec:summary}
Combining everything described above, we built a working prototype of the low latency video transmission system and demonstrated the remote car operation over commercially available LTE. 

\begin{figure}[t]%
\centering
\includegraphics[
width=\columnwidth
]%
{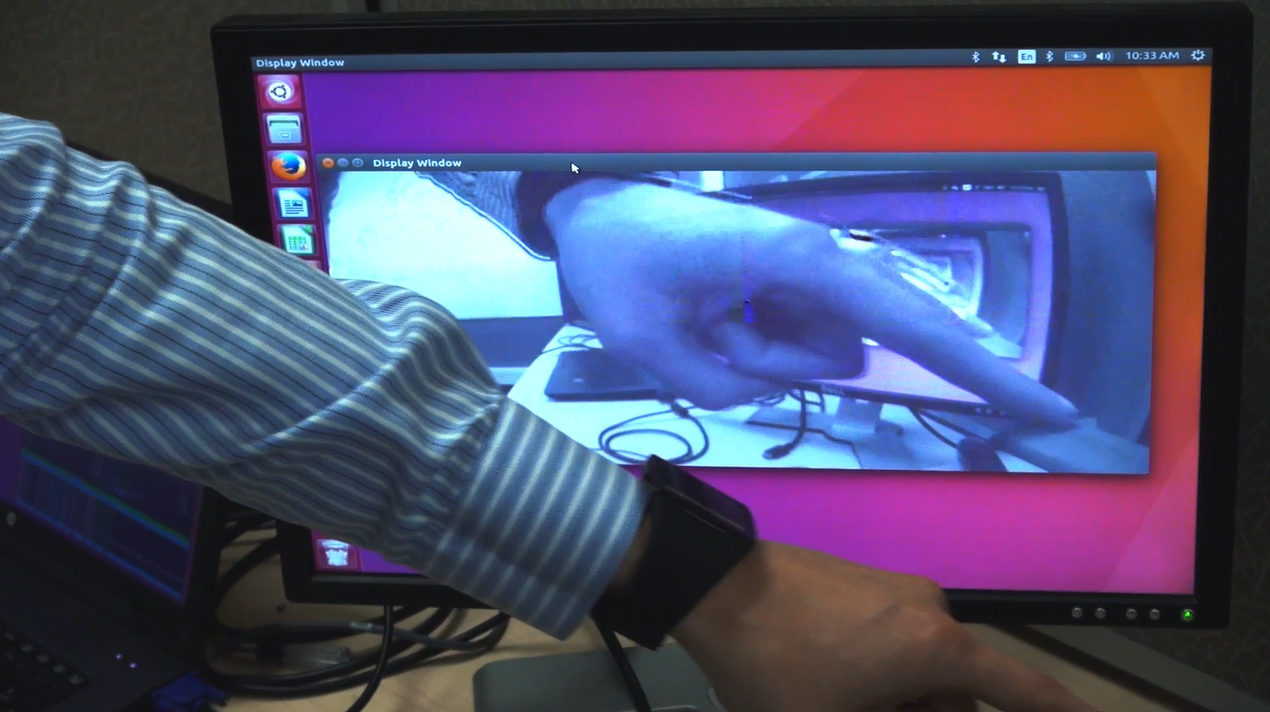}%
\caption{Video end-to-end latency measurement.}
\label{Fig:ete}%
\end{figure}

\begin{figure}[t]%
\centering
\includegraphics[
width=\columnwidth
]%
{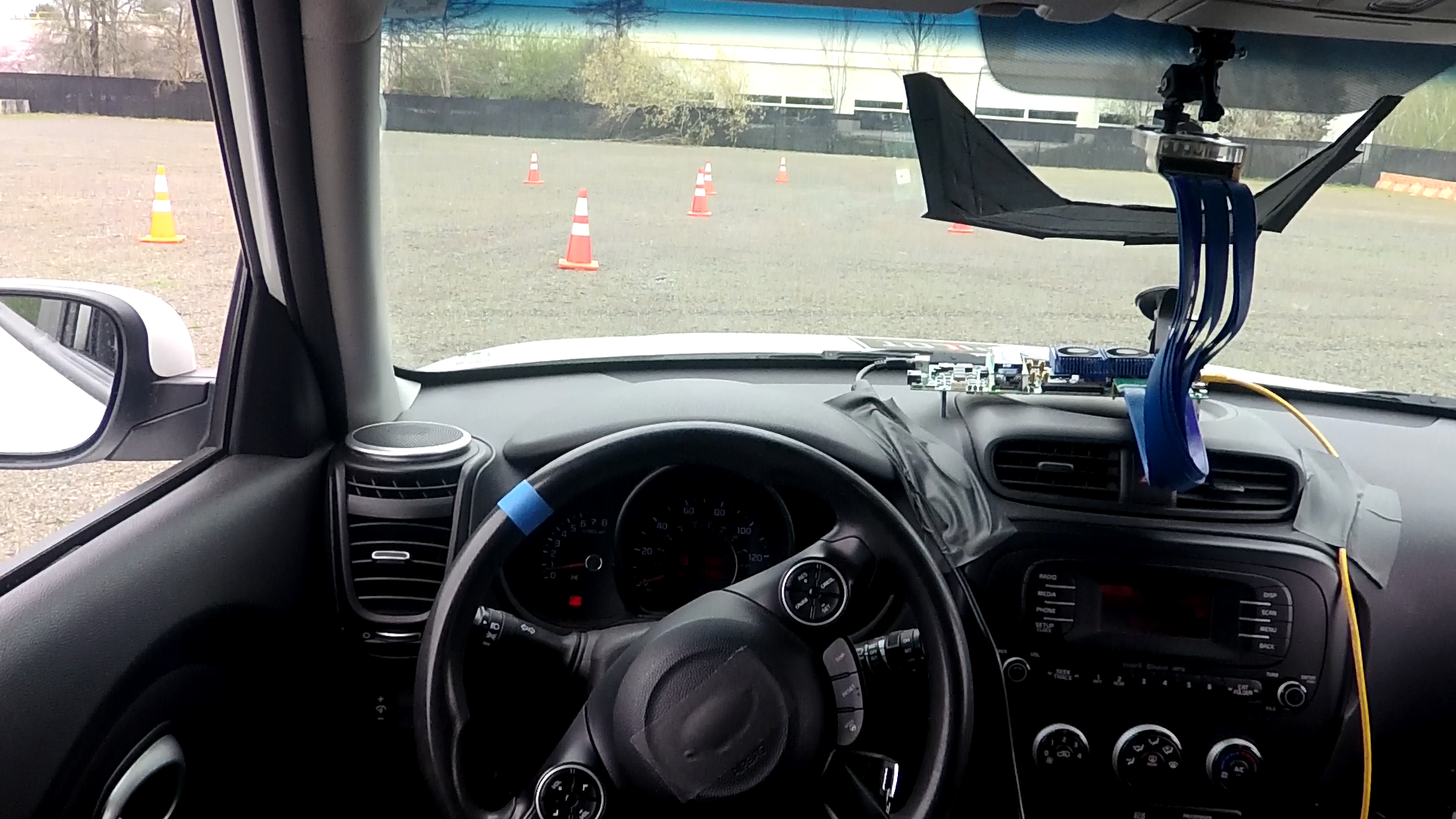}%
\caption{Remote driving test setup inside the car.}
\label{Fig:rd_inside}%
\end{figure}

\begin{figure}[t]%
\centering
\includegraphics[
width=\columnwidth
]%
{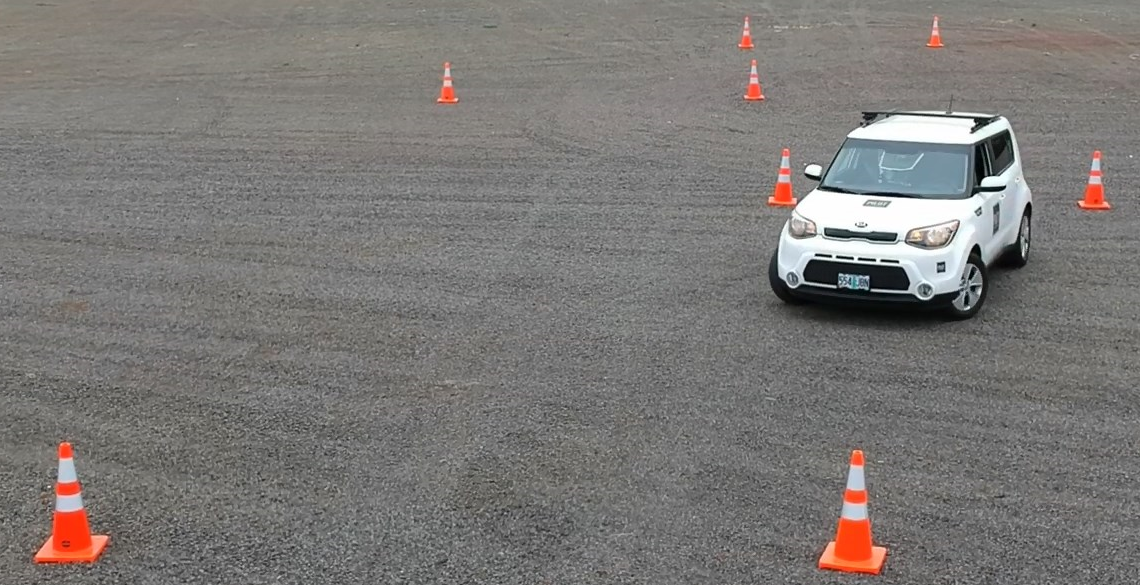}%
\caption{Remote driving test field.}
\label{Fig:rd_field}%
\end{figure}

With the prototype it was possible to achieve the following parameters: $46$ms average network roundtrip time and $90$ms average one-way video end-to-end latency measured with a standalone $60$fps camera. The average network roundtrip time was measured directly by time stamps, and video end-to-end video latency was measured as the amount of time from any event that happens in front of the camera to the event displayed by a remote screen, see Figure \ref{Fig:ete}. The measurement was performed by counting the number of frames in the video between $2$ events. As the camera frame rate was $60$fps, the granularity of the measurement was $16.67$ms. With the parameters obtained, we were able to operate a real car remotely demonstrated by precise steering between the cones at the company’s test field as shown in Figures see \ref{Fig:rd_inside} and \ref{Fig:rd_field}.

\begin{figure}[t]%
\centering
\scalebox{1}[0.6]{\includegraphics[
width=\columnwidth
]%
{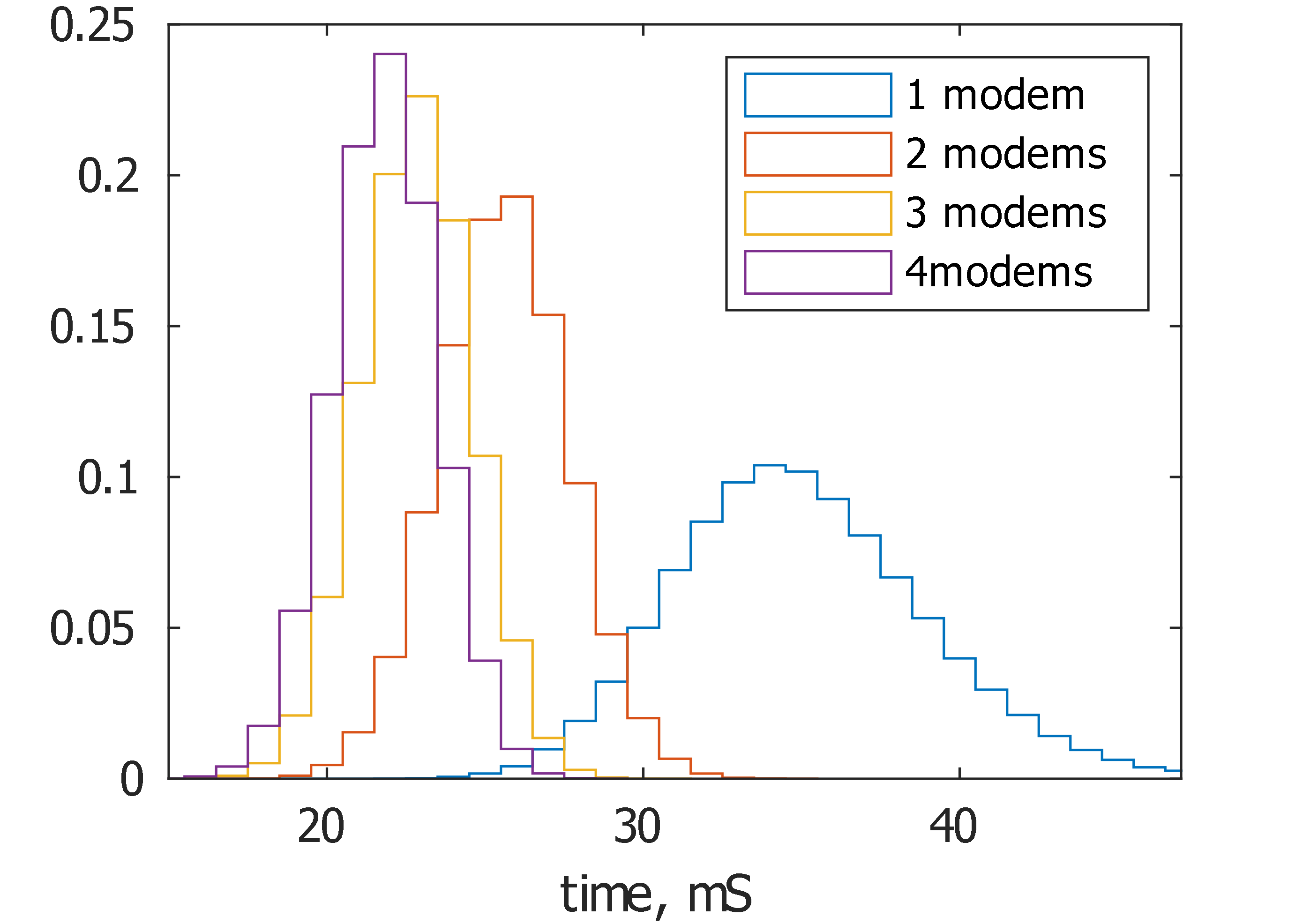}}%
\caption{Reconstructed slice latency distribution as a function of number of modems used.}
\label{Fig:last_his}%
\end{figure}

To achieve these latency numbers, we used two most impactful features:

1. At the network layer, we introduced multiple links by use of more than $1$ LTE modem. It was done not to increase the data rate, but to decorrelate the packet delays, so in combination with the rate control and scheduling, the principle of  network layer coding made possible to lower the latency. As increasing the number of modems increases the cost of the system, we don’t recommend to make latency reduction schemes using more than 4 modems. While having $2$ modems is the must, adding the $3$rd modem gives very small latency gain in average, latency gains of adding extra modems after $4$th would become marginal, as one can see in Figure \ref{Fig:last_his}, the average latency is changed by just $1$ms. It is important to note that adding of extra links changes the variance of the reconstructed slice delay: the more links we have, the lower is the variance. So, as the $4$th modem does not change the average latency, but it makes the latency closer to constant. 

2. At the application layer, we introduced slice-by-slice video processing by splitting the whole frame into smaller parts and processing them independently, also changing compression parameters on the fly. To make it happen, we had to build a proprietary hardware as there were no components available in the market. The hardware implementation option was chosen not only because of the real time operation requirements as video compression by itself can be efficiently implemented in software using latest SSE instructions, but because of the high data rates of the video at $60$fps. An HD $60$fps sensor would generate an uncompressed stream at $3$GBps rate, so even $1$Gbps ethernet that PCs have today is not enough to transmit the data. If compression used, the rate is lowered, but the latency grows, because the video stream needs to be transcoded in accordance with the required data rate at the network layer and has to be encoded using intra-refresh for streaming, but cameras available in the market use frame-by-frame compression that is another source of the latency and don’t generate streams within intra-refresh. Using hardware, we were able to read data from the sensor slice-by-slice and compress it providing data packets to the networking module for transmission directly. 

However, besides these two features we believe are most impactful, there is still room for the end-to-end latency improvements. For example, frame buffers at the display side. The same way we presented with tightly coupling of the sensor and the compression can be used at the display side to let data remain in a form of small blocks at the granularity lower than one frame up to the very last point of the system that is a display matrix, so display manufacturers may consider this option for their new products in the future.

\section*{ACKNOWLEDGMENT}
The authors would like to thank Rihards Gailums, CEO of PILOT Automotive Labs for helping with the vehicle used in remote driving demonstration and testing.

\section*{Appendix}

The final implementation of the networking module depicted in Figure \ref{Fig:networking} that was used in remote driving is released as an open source Intel\textregistered  Low Latency Multi Link Network Coding Library under Apache License 2.0, so we welcome researchers to access it at GitHub \cite{LLMLNCL}.

\bibliographystyle{IEEEtran}
\bibliography{remote_driving_bib}

% Generated by IEEEtran.bst, version: 1.12 (2007/01/11)
\begin{thebibliography}{1}
\providecommand{\url}[1]{#1}
\csname url@samestyle\endcsname
\providecommand{\newblock}{\relax}
\providecommand{\bibinfo}[2]{#2}
\providecommand{\BIBentrySTDinterwordspacing}{\spaceskip=0pt\relax}
\providecommand{\BIBentryALTinterwordstretchfactor}{4}
\providecommand{\BIBentryALTinterwordspacing}{\spaceskip=\fontdimen2\font plus
\BIBentryALTinterwordstretchfactor\fontdimen3\font minus
  \fontdimen4\font\relax}
\providecommand{\BIBforeignlanguage}[2]{{%
\expandafter\ifx\csname l@#1\endcsname\relax
\typeout{** WARNING: IEEEtran.bst: No hyphenation pattern has been}%
\typeout{** loaded for the language `#1'. Using the pattern for}%
\typeout{** the default language instead.}%
\else
\language=\csname l@#1\endcsname
\fi
#2}}
\providecommand{\BIBdecl}{\relax}
\BIBdecl

\bibitem{Kang18}
L.~Kang, W.~Zhao, B.~Qi, and S.~Banerjee, ``Augmenting self-driving with remote
  control: challenges and directions,'' in \emph{Proceedings of the 19th
  International Workshop on Mobile Computing Systems and Applications
  (HotMobile '18)}, Feb. 2018, pp. 19 -- 24.

\bibitem{Luck06}
J.~Luck, P.~McDermott, L.~Allender, and D.~Russell, ``An investigation of real
  world control of robotic assets under communication latency,'' in
  \emph{Proceedings of the 1st ACM SIGCHI/SIGART conference on Human-robot
  interaction (HIR’06))}, March 2006, pp. 202 -- 209.

\bibitem{Kabatiansky05}
G.~Kabatiansky, E.~Krouk, and S.~Semenov, \emph{Coding of Messages at the
  Transport Layer of the Data Network}.\hskip 1em plus 0.5em minus 0.4em\relax
  Wiley, 2005.

\bibitem{Luby07}
M.~Luby, A.~Shokrollahi, M.~Watson, and T.~Stockhammer, ``Raptor forward error
  correction scheme for object delivery,'' RFC 5053, Tech. Rep., 2007.

\bibitem{Taylor}
\BIBentryALTinterwordspacing
C.~Taylor. Fast {GF(256)} {Cauchy} {MDS} {Block} {Erasure} {CoDec} in {C}.
  [Online]. Available: \url{https://github.com/catid/cm256}
\BIBentrySTDinterwordspacing

\bibitem{Wiegand03}
T.~Wiegand, G.~Sullivan, G.~Bjontegaard, and A.~Luthra, ``Overview of the
  h.264/avc video coding standard,'' \emph{IEEE Transactions on Circuits and
  Systems for Video Technology}, vol.~13, no.~7, pp. 560--576, July 2003.

\bibitem{FFMPEG}
\BIBentryALTinterwordspacing
FFMPEG. A complete, cross-platform solution to record, convert and stream audio
  and video. [Online]. Available: \url{https://www.ffmpeg.org/}
\BIBentrySTDinterwordspacing

\bibitem{LLMLNCL}
\BIBentryALTinterwordspacing
A.~Belogolovy, E.~Stupachenko, and M.~Beacle. "intel(r) low latency multi link
  network coding library". [Online]. Available:
  \url{https://github.com/IntelLabs/LLMLNCL/}
\BIBentrySTDinterwordspacing

\end{thebibliography}

\end{document}